\documentclass[12pt]{article}
\usepackage{enumerate}
\usepackage{amsfonts}
\usepackage{amsmath}
\usepackage{amssymb}
\usepackage{amsthm}

\def\H{\mathcal{H}}

\def\S{\mathfrak{S}}

\def\T{\mathfrak{T}}

\newcommand{\rank}{\mathrm{rank}}

\newcommand{\Tr}{\mathrm{Tr}}

\newcommand{\shs}{\hspace{1pt}}

\newcounter{defin}  \newcounter{lemma}  \newcounter{theorem}
\newcounter{property} \newcounter{corol}  \newcounter{remark} \newcounter{example}

\newenvironment{theorem}{\par\refstepcounter{theorem}
     \textbf{Theorem \thetheorem.}\ }{\rm\par}

\newenvironment{corollary}{\par\refstepcounter{corol}
     \textbf{Corollary \thecorol.} }{\rm\par}

\begin{document}

\title{Simple characterization of positive linear maps preserving continuity of the von Neumann entropy}\author{M.E. Shirokov\footnote{Steklov Mathematical Institute, RAS, Moscow, email:msh@mi.ras.ru}}
\date{}
\maketitle

\begin{abstract}
We show that a positive linear map preserves local continuity (convergence) of the entropy if and only if it preserves finiteness of the entropy, i.e. transforms operators with finite entropy to operators with finite entropy. The last property is equivalent to the boundedness of the output entropy of a map on the set of pure states.
\end{abstract}

\section{Preliminaries}

Linear completely positive maps between Banach spaces of trace-class operators play a basic role in description of evolution of quantum systems and  of measurements of their parameters. Such maps are called \emph{quantum channels} (respectively, \emph{quantum operations}) if they preserve (respectively, don't increase) the trace of any input positive operator~\cite{H-SCI}.

In analysis of information properties of infinite-dimensional quantum channels and operations the question about continuity of their output entropy as a function of input states naturally appears~\cite{OE}. In this note we give a characterization of positive maps (in particular, quantum channels and operations) for which continuity of the  entropy on a set of input operators (states) implies continuity of the output entropy on this set.

\pagebreak

Let $\H$~be a separable Hilbert space,
$\mathfrak{T}(\H)$ the Banach space of trace class operators on~$\H$, $\mathfrak{T}_{+}(\H)$~the cone of positive operators in~$\mathfrak{T}(\H)$ and
$\mathfrak{S}(\H)$ the set of \emph{quantum states}~-- operators in ~$\mathfrak{T}_{+}(\H)$ with unit trace~\cite{H-SCI}.

\smallskip

The \emph{von Neumann entropy} of a quantum state
$\rho \in \mathfrak{S}(\H)$ is  defined by the formula
$H(\rho)=\operatorname{Tr}\eta(\rho)$, where  $\eta(x)=-x\log x$ for $x>0$
and $\eta(0)=0$. It is a concave lower semicontinuous function on the set~$\mathfrak{S}(\H)$ taking values in~$[0,+\infty]$ (see~\cite{H-SCI,L-2,W}).
The homogeneous extension of the von Neumann entropy to the cone~$\mathfrak{T}_+(\H)$ is given by the formula:
$$
H(\rho)\doteq[\Tr\rho]H\biggl(\frac{\rho}{\Tr\rho}\biggr)=
\Tr\eta(\rho)-\eta(\Tr\rho).
$$
By using theorem 11.10 in \cite{N&Ch} and a simple approximation it
is easy to obtain the following inequality
\begin{equation}\label{w-k-ineq}
H\!\left(\sum_{k}p_{k}\rho_{k}\right)\leq
\sum_{k}p_{k}H(\rho_{k})+S(\{p_{k}\}_k),
\end{equation}
valid for any finite or countable collection  $\{\rho_{k}\}$ of positive operators in
the unit ball of $\mathfrak{T}(\mathcal{H})$ and any probability distribution
$\{p_k\}$, where
\begin{equation}\label{sh-ent}
S(\{p_k\})\doteq\sum_k\eta(p_k)-\eta\left(\sum_kp_k\right)
\end{equation}
is the homogeneous extension of the Shannon entropy to the positive cone  of $\ell_1$.
Inequality (\ref{w-k-ineq})
implies the following one
\begin{equation}\label{w-k-ineq+}
H\!\left(\sum_{k}\rho_{k}\right)\leq\sum_{k}H(\rho_{k})+S\!\left(\left\{\mathrm{Tr}\rho_{k}\right\}_{k}\right),
\end{equation}
valid for any finite or countable collection  $\{\rho_{k}\}$ of positive operators in $\mathfrak{T}(\mathcal{H})$ with finite
$\sum_{k}\!\mathrm{Tr}\rho_{k}$. If $\mathrm{supp}\rho_k \perp\mathrm{supp}\rho_j$ for all $k\neq j$ then $"="$ holds in (\ref{w-k-ineq+}).\footnote{The support $\,\mathrm{supp}\rho\,$ of a positive operator $\rho$ is the orthogonal complement to its kernel.}
\vspace{5pt}

The \emph{quantum relative entropy} for positive trace class operators  $\rho$ and
$\sigma$  can be defined as
\begin{equation}\label{re-def}
H(\rho\shs\|\shs\sigma)=\sum_i\langle
i|\,\rho\log\rho-\rho\log\sigma+\sigma-\rho\,|i\rangle,
\end{equation}
where $\{|i\rangle\}$ is the orthonormal basis of
eigenvectors of the state $\rho$ and it is assumed that
$H(\rho\shs\|\shs\sigma)=+\infty$ if $\,\mathrm{supp}\rho\shs$ is not
contained in $\shs\mathrm{supp}\shs\sigma$ \cite{L-2}.

\section{PCE-property and its characterization}

For an arbitrary positive linear map
$\Phi\colon\mathfrak{T}(\H_A)\to\mathfrak{T}(\H_B)$ the output entropy $H_{\Phi}(\rho)\doteq H(\Phi(\rho))$~is a concave lower semicontinuous function on the cone~$\mathfrak{T}_+(\H_A)$ taking values in~$[0,+\infty]$.

The class of positive maps characterised by continuity of the output entropy $H_{\Phi}(\rho)$ on any subset of the cone~$\mathfrak{T}_+(\H_A)$ on which the input entropy~$H(\rho)$ is continuous was considered in~\cite{OE}. This property can be called
\emph{preserving  continuity of the entropy (PCE)} under action of a map~$\Phi$. A simple characterization of the  PCE\nobreakdash-\hspace{0pt}property is given by the following theorem (which is a strengthened version of Theorem 2 in \cite{OE}).\smallskip

\begin{theorem}\label{main}
\emph{Let $\,\Phi$~be a  positive linear map from~$\mathfrak{T}(\H_A)$ into~$\mathfrak{T}(\H_B)$.
The following properties are equivalent:}\smallskip

{\rm(i)} \emph{$\,\Phi$~preserves continuity of the entropy, i.e.
$$
\lim_{n\to\infty}H(\rho_n)=H(\rho_0)<+\infty\quad\Rightarrow\quad\lim_{n\to\infty}H_{\Phi}(\rho_n)=H_{\Phi}(\rho_0)<+\infty
$$
for any sequence $\{\rho_n\}\subset\mathfrak{T}_+(\H_A)$ converging to an operator~$\rho_0$;}\smallskip

{\rm(ii)} \emph{$\,\Phi$~preserves finiteness of the entropy, i.e.
$$
H(\rho)<+\infty\quad\Rightarrow\quad H_{\Phi}(\rho)<+\infty
$$
for any state $\rho\in\mathfrak{S}(\H_A)$;}\smallskip

{\rm(iii)} \emph{the function~$H_{\Phi}(\rho)$ is bounded on the set $\,\mathrm{ext}\mathfrak{S}(\H_A)$ of pure states.}\medskip
\end{theorem}

\emph{Proof.} We may assume that $\Phi$ does not increase a trace of any positive operator.

It is clear that \rm(i) implies \rm(ii).

By using inequality (\ref{w-k-ineq}) and the spectral decomposition $\rho=\sum_ip_i |\varphi_i\rangle\langle\varphi_i|$ we obtain
\begin{equation}\label{H-ub}
H_{\Phi}(\rho)\leq \sum_ip_i H_{\Phi}(|\varphi_i\rangle\langle\varphi_i|)+H(\rho)\leq C\|\rho\|_1+H(\rho),
\end{equation}
where $\,C\doteq\sup_{\rho\in\mathrm{ext}\mathfrak{S}(\H_A)}H_{\Phi}(\rho)$. This shows that \rm(iii) implies \rm(ii).\smallskip

To prove the implication $\,\mathrm{(ii)\Rightarrow(iii)}$ assume that for each natural $n$ there exists a pure state $\rho_n$ such that $H_{\Phi}(\rho_n)\geq 2^n$. It follows from (\ref{w-k-ineq}) that the state $\rho_0=\sum_{n\in\mathbb{N}}2^{-n}\rho_n$ has finite entropy, while the concavity of the function $H_{\Phi}$ implies $H_{\Phi}(\rho_0)\geq \sum_{n\in\mathbb{N}}2^{-n}H_{\Phi}(\rho_n)=+\infty$.\smallskip

To prove the nontrivial implication $\,\mathrm{(iii)\Rightarrow(i)}$ it suffices to show that \rm(iii) implies continuity of the function $H_{\Phi}$ on the set $\mathrm{ext}\S(\H_A)$ and to apply Theorem 2 in \cite{OE}. But we will give an independent proof of this implication based on the recently established monotonicity of the relative entropy under  positive linear maps \cite{MRE-2}.

Show first that \rm(iii) implies continuity of the function $H_{\Phi}$ on the set
\begin{equation}\label{s-k}
\mathfrak{S}_k(\H_A)=
\left\{\rho\in\mathfrak{S}(\H_A)\mid \operatorname{rank}\rho\leq k\right\}
\end{equation}
for any  $\,k$ by using Winter's modification of the Alicki-Fannes method \cite{A&F,W-CB}.

Let $\rho$ and $\sigma$ be different states in $\mathfrak{S}_k(\H_A)$ and $\,\varepsilon=\frac{1}{2}\|\rho-\sigma\|_1$. Following \cite{W-CB} introduce the state
$\,\omega^{*}=(1+\varepsilon)^{-1}(\rho+[\shs\sigma-\rho\shs]_+)$. Then
\begin{equation*}
\frac{1}{1+\varepsilon}\,\rho+\frac{\varepsilon}{1+\varepsilon}\,\tau_-=\omega^{*}=
\frac{1}{1+\varepsilon}\,\sigma+\frac{\varepsilon}{1+\varepsilon}\,\tau_+,
\end{equation*}
where $\,\tau_+=\varepsilon^{-1}[\shs\rho-\sigma\shs]_+\,$
and
$\,\tau_-=\varepsilon^{-1}[\shs\rho-\sigma\shs]_-\,$
are states in $\S(\H_{A})$. Since $\rho$ and $\sigma$ are different states in $\S_k(\H_{A})$, the rank of the states
$\tau_+$ and $\tau_-$ does not exceed $2k-1$. It follows from (\ref{H-ub}) that
\begin{equation}\label{H-tau}
H_{\Phi}(\tau_\pm)\leq C+\log(2k-1).
\end{equation}

The concavity of the entropy and inequality (\ref{w-k-ineq}) imply
\begin{equation*}
p
H_{\Phi}(\varrho)+(1-p)H_{\Phi}(\varsigma)\leq H_{\Phi}\!\left(p\varrho+(1-p)\varsigma\right)\leq p
H_{\Phi}(\varrho)+(1-p)H_{\Phi}(\varsigma)+h_2(p)
\end{equation*}
for any $p\in(0,1)$ and any states $\varrho$ and $\varsigma$, where $h_2(p)=\eta(p)+\eta(1-p)$. \smallskip

By applying this double inequality to the above convex decompositions of $\,\omega_{*}$ we obtain
$$
(1-p)\left[H_{\Phi}(\rho)-H_{\Phi}(\sigma)\right]\leq p
\left[H_{\Phi}(\tau_+)
-H_{\Phi}(\tau_-)\right]+
h_2(p)
$$
and
$$
(1-p)\left[H_{\Phi}(\sigma)-H_{\Phi}(\rho)\right]\leq p
\left[H_{\Phi}(\tau_-)-H_{\Phi}(\tau_+)\right]+\shs h_2(p),
$$
where $p=\frac{\varepsilon}{1+\varepsilon}$.

These inequalities and  upper bound (\ref{H-tau}) show that
$$
|H_{\Phi}(\rho)-
H_{\Phi}(\sigma)|\leq \varepsilon (C+\log(2k-1))+(1+\varepsilon)h_2\!\left(\frac{\varepsilon}{1+\varepsilon}\right).
$$
This implies (uniform) continuity of the  function $H_{\Phi}(\rho)$ on the set
$\mathfrak{S}_k(\H_A)$.\smallskip

We will prove \rm(i) by using the approximation technique proposed in \cite{SSP}. By Proposition 3 in \cite{SSP}  any concave lower semicontinuous nonnegative function $f$ on $\S(\H)$ is a pointwise limit of the nondecreasing sequence $\{\hat{f}_k\}$ of concave lower semicontinuous nonnegative functions on $\S(\H)$ defined as follows
 \begin{equation}\label{f-k-def}
\hat{f}_k(\rho)=\sup_{\{\pi_{i},\rho_{i}\}\in\mathcal{P}_{k}(\rho)}
\sum_{i}\pi_{i}f(\rho_{i}),\qquad \rho\in
\S(\mathcal{H}),
\end{equation}
where $\mathcal{P}_{k}(\rho)$ is the set of all countable ensembles of states in $\S_k(\H)$ with the average state $\rho$. If the function $f$ is continuous and bounded on the set $\S_k(\H)$ for any natural $k$ then all the functions $\hat{f}_k$ are continuous on $\S(\mathcal{H})$.\footnote{This is a corollary of the strong stability of the set $\S(\H)$ \cite{SSP}.} So, in this case Dini's lemma implies the following criterion for local continuity of the function $f$:  \emph{the function $f$ is continuous on a compact subset of $\S(\H)$ if and only if  the sequence $\{\hat{f}_k\}$ uniformly converges to the function $f$ on this subset.}

This criterion gives a powerful method for  analysis of continuity of the von Neumann entropy (described in \cite{SSP}), since in the case $f=H$
the difference
$\,\Delta^H_{k}(\rho)=H(\rho)-\hat{H}_{k}(\rho)\,$ between the von Neumann
entropy and its $k$-approximator $\hat{H}_{k}$ is expressed via the quantum relative entropy as follows
\begin{equation}\label{Delta}
\Delta^H_{k}(\rho)=\inf_{\{\pi_{i},\rho_{i}\}\in\mathcal{P}_{k}(\rho)}
\sum_{i}\pi_{i}H(\rho_{i}\|\rho). \smallskip
\end{equation}

Since we have shown that \rm(iii) implies continuity and boundedness of the function $H_{\Phi}$ on the set $\S_k(\H_A)$ for each $k$, we can
use the above criterion of local continuity in the case $f=H_{\Phi}$. The concavity of the function $\eta(x)=-x\log x$ implies that
\begin{equation}\label{Delta+}
\Delta^{H_{\Phi}}_{k}(\rho)\doteq H_{\Phi}(\rho)-[\widehat{H_{\Phi}}]_{k}(\rho)\leq\inf_{\{\pi_{i},\rho_{i}\}\in\mathcal{P}_{k}(\rho)}
\sum_{i}\pi_{i}H(\Phi(\rho_{i})\|\Phi(\rho)). \smallskip
\end{equation}

By the monotonicity  of the relative entropy (defined by formula (\ref{re-def})) under trace-non-increasing positive linear maps (see the Appendix) it follows from (\ref{Delta}) and (\ref{Delta+}) that
$$
\Delta^{H_{\Phi}}_{k}(\rho)\leq\Delta^H_{k}(\rho)\quad\textrm{ for all }\rho\in\S(\H_A).
$$
Hence, \emph{uniform convergence of the sequence $\,\{\hat{H}_{k}\}\,$ to the function $H$ on any subset of $\,\S(\H_A)$  implies
uniform convergence of the sequence $\,\{[\widehat{H_{\Phi}}]_{k}\}\,$ to the function $H_{\Phi}$ on this subset.} By combining this observation with the above criterion of local continuity in the cases  $f=H$ and $f=H_{\Phi}$ we obtain the implication in \rm(i)
for any sequence $\{\rho_n\}$ of \emph{states} converging to a state ~$\rho_0$. Since the functions $H$ and $H_{\Phi}$ are homogeneous (of degree 1) to prove the general form of \rm(i) it suffices to show that $\,\lim_{n}H_{\Phi}(\rho_n)=0\,$ for any sequence $\{\rho_n\}$ in $\T_+(\H_A)$ converging to the zero operator such that $\,\lim_{n}H(\rho_n)=0$. This can be done by using (\ref{H-ub}). $\square$
\smallskip

The arguments from the proof of Theorem \ref{main} imply the following\smallskip

\begin{corollary}\label{main-c1} \emph{If $H_{\Phi}(\rho)\leq C<+\infty$ for any pure state $\rho\in\mathfrak{S}(\H_A)$ then
the function~$H_{\Phi}$ is uniformly continuous on the set
$\,\mathfrak{S}_k(\H_A)$ defined in (\ref{s-k}) for any natural $\,k$ and
$$
|H_{\Phi}(\rho)-
H_{\Phi}(\sigma)|\leq \varepsilon (C+\log(\rank\rho+\rank\sigma-1))+(1+\varepsilon)h_2\!\left(\frac{\varepsilon}{1+\varepsilon}\right)
$$
for any finite rank states  $\,\rho$ and $\,\sigma$ in $\,\S(\H_A)$, where $\;\varepsilon=\frac{1}{2}\|\shs\rho-\sigma\|_1$.}
\end{corollary}\smallskip

The simplest  PCE-maps are maps with finite-dimensional output and unitary transformations, i.e. maps of the form $\Phi(\rho)=U\rho U^*$, where  $U$~is an isometry from~$\H_A$ into~$\H_B$. More interesting examples are quantum channels and operations with finite Choi rank. The PCE-channel with infinite input and output spaces and infinite Choi rank is discribed  after Proposition 2 in \cite{HF}.\smallskip

Further results in this direction, in particular, the classification of\break PCE-channels and quantitative continuity analysis of their output entropy, are obtained in \cite{PFE}.
\bigskip

\section*{Appendix: On monotonicity of the relative entropy under trace-non-increasing positive maps}

Recently Muller-Hermes and Reeb established (essentially basing on Beigi's results \cite{MRE-1}) the following fundamental property.\smallskip

\begin{theorem}\label{m} \cite{MRE-2} \emph{If $\,\Phi$ is a positive trace-preserving linear map then}
\begin{equation}\label{m-ineq}
H(\Phi(\rho)\|\Phi(\sigma))\leq H(\rho\|\sigma)\quad\textit{ for any states }\rho\textit{ and }\sigma.
\end{equation}
\end{theorem}

Muller-Hermes and Reeb mentioned in \cite{MRE-2} that this result is not generalized to trace-non-increasing positive linear maps until we use for all positive trace-class operators the same definition of the relative entropy as for quantum states. But using Lindblad's definition (\ref{re-def}) of the relative entropy it is easy to obtain such generalization. \smallskip

\begin{corollary}\label{m-c} \emph{If the relative entropy is defined by formula (\ref{re-def}) then (\ref{m-ineq}) is  valid for any
trace-non-increasing positive linear map $\Phi$.}
\end{corollary}\smallskip

\emph{Proof.} Consider the  trace-preserving positive  map $\Phi'(\rho)=\Phi(\rho)\oplus \Psi(\rho)$ from $\T(\H_A)$ to $\T(\H_B\oplus\H_C)$, where $\Psi(\rho)=[\Tr(\rho-\Phi(\rho))]\tau\,$ is a positive linear map from $\T(\H_A)$ to $\T(\H_C)$ determined by a given state $\tau$ in $\S(\H_C)$. Then
$$
H(\Phi'(\rho)\|\Phi'(\sigma))=H(\Phi(\rho)\|\Phi(\sigma))+H(\Psi(\rho)\|\Psi(\sigma)).
$$
The nonnegativity of the relative entropy  defined by formula (\ref{re-def}) and Theorem \ref{m} imply
$$
H(\Phi(\rho)\|\Phi(\sigma))\leq H(\Phi'(\rho)\|\Phi'(\sigma))\leq H(\rho\|\sigma)
$$
for any states $\rho$  and $\sigma$. $\square$

\bigskip

I am grateful to A.S.Holevo and G.G.Amosov for useful discussion.

\end{document}